\documentclass[useAMS,usenatbib,times]{mn2e}

\usepackage{graphicx}

\newcommand{\Msol}{\mbox{$M_{\sun}$}}

\newcommand{\Mpch}{\mbox{$h^{-1}\mathrm{Mpc}$}}
\newcommand{\Map}{\mbox{$M_{\mathrm{ap}}$}}

\title[Optimizing weak lensing mass estimates for cluster profile uncertainty]{Optimizing weak lensing mass estimates for cluster profile uncertainty}
\author[D. Gruen, G.~M. Bernstein, T.~Y. Lam, S. Seitz]{D. Gruen$^{1,2}$\thanks{E-mail:
dgru@sas.upenn.edu (DG)}, G.M. Bernstein$^{1}$, T.Y. Lam$^{3}$, S. Seitz$^{2,4}$\\
$^{1}$Department of Physics and Astronomy, 209 South 33rd Street, University of Pennsylvania, Philadelphia, PA 19104, USA\\
$^{2}$University Observatory Munich, Scheinerstrasse 1, 81679 Munich, Germany\\
$^{3}$Institute for the Physics and Mathematics of the Universe, University of Tokyo, Kashiwa, Chiba 277-8583, Japan\\
$^{4}$Max Planck Institute for Extraterrestrial Physics, Giessenbachstrasse, 85748 Garching, Germany}
\begin{document}

\date{}

\pagerange{\pageref{firstpage}--\pageref{lastpage}} \pubyear{2011}

\maketitle

\label{firstpage}

\begin{abstract}
Weak lensing measurements of cluster masses are necessary for calibrating mass-observable relations (MORs) to investigate the growth of structure and the properties of dark energy. However, the measured cluster shear signal varies at fixed mass $M_{200m}$ due to inherent ellipticity of background galaxies, intervening structures along the line of sight, and variations in the cluster structure due to scatter in concentrations, asphericity and substructure. We use $N$-body simulated halos to derive and evaluate a weak lensing circular aperture mass measurement $\Map$ that minimizes the mass estimate variance $\langle(\Map-M_{200m})^2\rangle$ in the presence of all these forms of variability. Depending on halo mass and observational conditions, the resulting mass estimator improves on $\Map$ filters optimized for circular NFW-profile clusters in the presence of uncorrelated large scale structure (LSS) about as much as the latter improve on an estimator that only minimizes the influence of shape noise. Optimizing for uncorrelated LSS while ignoring the variation of internal cluster structure puts too much weight on the profile near the cores of halos, and under some circumstances can even be worse than not accounting for LSS at all. We briefly discuss the impact of variability in cluster structure and correlated structures on the design and performance of weak lensing surveys intended to calibrate cluster MORs.
\end{abstract}


\section[]{Introduction}

Counts of dark matter halos and their clustering properties as a function of halo mass and redshift are a useful tool of observational cosmology, allowing both the determination of parameters of $\Lambda$CDM cosmology \citep{1998ApJ...508..483W,2001ApJ...560L.111H} and constraints on the physics of Dark Energy \citep{PhysRevLett.88.231301} in a way that is complementary to other probes \citep{2007NJPh....9..446T,2009PhRvD..80f3532C}. Upcoming and ongoing surveys such as the Dark Energy Survey (DES) \citep{2005astro.ph.10346T}, or the Atacama Cosmology Telescope (ACT) and South Pole Telescope (SPT) surveys will make such measurements of the non-linear growth of structure.

The detection of clusters and measurement of their masses can be performed using various signals, such as galaxy density and the abundance of red sequence galaxies \citep{1999ApJ...517...78K,2000AJ....120.2148G}, the Sunyaev-Zel'dovich (SZ) effect \citep{1972CoASP...4..173S,2001ApJ...553..545H,2003PhRvD..68h3506B} or measurements of the cluster X-ray emission \citep[e.g.][]{2004A&A...425..367B,2009MNRAS.397..577S,2010arXiv1007.1916P}. In principle, SZ or X-Ray surveys can use a self-calibrated mass-observable relation (MOR) \citep{2003PhRvD..67h1304H,2004ApJ...613...41M}. However, constraints on cosmological parameters can be greatly improved if the MOR can be determined independently. Weak gravitiational lensing has therefore become an important tool for finding the MOR of these methods \citep[e.g.][]{2002MNRAS.334L..11A,2010ApJ...721..875O,2011ApJ...726...48H}, since the lensing signal is most closely related to the true mass of a cluster. As massive objects leave their imprint on space-time, their presence is revealed by the weak distortion of images of galaxies behind them at small angular separations, allowing for reconstruction of their masses, which can then be used to calibrate other mass estimators. A recently developed alternative weak lensing approach is shear peak statistics, which can be used to constrain cosmology based on the lensing signal of the projected matter density and without the requirement for individual halo estimates of the virial mass \citep{2006PhRvD..73l3525M,2009ApJ...698L..33M,2010MNRAS.402.1049D}.

Halos can be detected with weak lensing for instance by applying inversion of the shear signal for generating convergence $\kappa$ maps \citep{1993ApJ...404..441K,1995A&A...294..411S,1995A&A...297..287S,1996A&A...305..383S,2001A&A...374..740S} or the application of an appropriate filter \citep[see Section~\ref{sec:mvf} and][]{1995ApJ...439L...1K,1996MNRAS.283..837S}, from which cluster candidates can be identified \citep[e.g.][]{2007A&A...462..875S,2007ApJ...669..714M}. Clusters detected in this way (or through baryonic signatures) can have mass estimates refined, 
depending on the strength of their lensing signal, by strong lensing mass profile reconstructions, by applying linear filters to the shear field, and/or by fitting a profile to the shear values observed. 

While strong lensing is complementary and only possible in relatively rare cases \citep[e.g.][]{1998AJ....116.1541A,2005A&A...437...39B,2006MNRAS.372.1425H}, the last two methods are generally applicable ways of using the lensing signal. 
We will assume that weak-lensing mass estimators operate on the azimuthally averaged tangential reduced shear profile, $g_t(\theta)$, which suffers from several sources of noise. For one thing, background galaxies are not intrinsically round, adding white noise to the shear measurement (\textbf{shape noise}) inversely proportional to the background galaxy density. Intrinsic alignments of background galaxies with respect to each other might introduce non-zero covariance between $g_t$ at different radii and must be treated carefully \citep{2006MNRAS.367..611M}, but we expect the intrinsic-alignment signal to be small compared to other forms of variance, so we will not consider it further.
Secondly, any \textbf{uncorrelated structures} along the line of sight between observer and background sources will introduce an additional weak lensing signal. While the expectation value of $g_t$ from uncorrelated structures is zero, its variance and covariance must be taken into account when designing an optimal method of weak lensing measurements increase the uncertainty of weak lensing masses \citep{2001A&A...370..743H,2003MNRAS.339.1155H,2010arXiv1011.1084H}. Thirdly, dark matter halos are more likely to occur in overdense regions, where the probability of the presence of additional halos is higher than elsewhere. This can cause a signal with non-zero expectation value and significant variance that disturbs weak lensing mass estimates \citep{2001ApJ...547..560M}.

The first two of these effects have been addressed in several ways. Shape noise is relatively well-controllable and deeper images with a larger number of background sources can at least in principle reduce it to a level that does not limit the intended scientific application, essentially allowing for a more complete and clean detection of less massive halos. Intrinsic alignments can be accounted for when redshift information is present \citep{2003A&A...398...23K,2003MNRAS.339..711H,2008A&A...488..829J}. The influence of shear covariance due to uncorrelated structures on mass estimates can be accounted for when constructing a filter to measure the amplitude of a known true shear profile \citep{2004PhRvD..70b3008D,2005A&A...442..851M}. 

What has not been accounted for, so far, is the variability of the cluster profile itself due to correlated halos and internal structure. The lensing signal of correlated halos cannot be discriminated from the cluster signal itself and, after appropriate calibration of the mass estimate, remains as a source of noise. This might be particularly severe as correlated halos align along a cosmic web. There is more to intrinsic variability, however, than correlated 
halos. Profile fitting methods and optimized linear-filter mass estimators assume the presence of a true shear profile, typically the characteristic shear signature of Navarro-Frenk-White (NFW) profile dark matter halos \citep{1996ApJ...462..563N,1996A&A...313..697B,2000ApJ...534...34W}. The average profile of dark matter halos as measured from simulations or with weak lensing is consistent with the NFW profile \citep{2006MNRAS.372..758M,2007arXiv0709.1159J}, although there are claims for small deviations in the core \citep{2006AJ....132.2685M} and at large radii \citep{2011MNRAS.tmp..809O}. What is certainly {\em not} true, however, is that {\em individual} halos follow spherical NFW profiles with concentrations that only depend on their masses. Asphericity of halos \citep{1992ApJ...399..405W} can lead to an overestimation of the mass of clusters aligned with the line of sight and an underestimation for those aligned perpendicular to it, especially when fitting NFW shear profiles to observed shear \citep{2004MNRAS.350.1038C,2007MNRAS.380..149C}. Linear filters, even those adapted to different mass ranges, will suffer from the width of the distribution of concentration parameters at a given mass \citep[cf.][for measurements of the concentration distributions]{2001MNRAS.321..559B}. Fitting for concentration as a second parameter, however, introduces 
the problem of degeneracy with the primary measurement of mass. Additionally, dark matter halos contain substructure, different amounts of baryonic matter and correlated structure outside the virial radius that induce variance in the actual profiles. When constructing a filter that yields a minimum variance mass estimate, the variance of true shear profiles could be taken into consideration as an additional component of the noise if there was an appropriate analytic model. The variety of effects involved, some of which are not entirely understood, makes this very difficult, however.

We subsume all profile variability due to the effects mentioned in the previous paragraph under the term \textbf{correlated structures} in what follows. We take them into account by using cosmological simulations of dark matter halos, which include all kinds of correlated structures. Since we know the true $M_{200m}$ masses of these simulated halos as well as the reduced shear profile $g_t(\theta)$ including all correlated structures, we can derive the linear function $\Map$ of $g_t$ that yields the minimum variance $\langle (\Map-M_{200m})^2\rangle$ in the lensing estimation of the cluster virial masses.
Thus, we drop the usual assumption of a universal shear profile, find the effect on cluster mass reconstruction, and re-optimize mass estimators taking the correlated structures into account. By taking into account not only shape noise and uncorrelated LSS, but also correlated structures, we give a full estimate of the scatter in lensing mass estimates.

The structure of this paper is as follows. In Section 2 we discuss our methodology for finding a minimum variance filter in the presence of realistic cluster profiles and uncorrelated LSS. In Section 3 we investigate the radial dependence of halo profile variability and calculate and analyze the minimum variance filter for the simulated halos. In the final Section 4 we discuss our results and implications for the design of weak lensing cluster filters in future surveys. The Appendix contains details of the halo model formalism used to calculate the covariance matrix of the LSS lensing signal.

\section[]{Methods}

\subsection[]{Simulations}

We use a simulation that is part of the one analysed by \citet{2010arXiv1011.1681B}. It was made using $1024^3$ dark matter particles of mass $6.98\cdot 10^{10} h^{-1}\Msol$ in a box of comoving size 1 Gpc$ h^{-1}$ with a parallelized Adaptive Refinement Tree algorithm \citep{1997ApJS..111...73K,2008arXiv0803.4343G}. It is simulated from redshift $z=60$ to $z=0$ with cosmological parameters $(\Omega_m,\Omega_b,\sigma_8,h,n)=(0.27,0.044,0.79,0.7,0.95)$ at an effective spacial resolution of $30$~kpc$ h^{-1}$. The same simulation was used in \citet{2008ApJ...688..709T}, where it was labeled L1000W. We use a snapshot at redshift $z_d=0.245$, in which 14,856 halos of $M_{200m}\geq 1\cdot 10^{14}\Msol$ were identified, 24 of which are heavier than $M_{200m}=1\cdot 10^{15}\Msol$. 

For the selected halos, the mass is integrated along the line of sight and the lensing signal is determined on a grid of approximately $40$ comoving kpc$/$pixel using the Born approximation as described in \citet[Section~3]{2010arXiv1011.1681B}, taking into account all matter within comoving $\pm200\Mpch$ along the line of sight and, transversely, in a square with comoving side length $20\Mpch$ centered on the cluster. We assume background sources to be at a constant redshift $z_s=1$, which at our $z_d=0.245$ scales gravitational shear by a factor of $0.76$ compared to sources at infinity. We extract from this tangential reduced shears in radial bins beginning at $r_0=1'$ and having a width $r_i\in[(1-\Delta)\cdot r_i,(1+\Delta)\cdot r_i]$ with $\Delta=0.1$. Shear inside $r=1'$ (approximately 160~kpc proper radius) is subject to resolution effects and biased low because of smoothing, which is why we discard it for our analysis. Since the total weight of the excluded region is low, our results do not change significantly if the signal inside $r=1'$ is included.

\subsection[]{Minimum Variance Filter}
\label{sec:mvf}

A useful approach of measuring halo masses is to use a filter $u(\btheta)$ that weights the projected mass density within an aperture \citep{1995ApJ...439L...1K,1996MNRAS.283..837S}. Let for all the following $g$, $\gamma$ and $\kappa$ be the azimuthally averaged $\bar{g}_t$, $\bar{\gamma}_t$ and $\bar{\kappa}$ at the respective radius. The aperture mass within an aperture $u$ is defined as
\begin{equation}
\Map = 2\pi \int d\theta \; \theta \; u(\theta) \kappa(\theta) \; .
\end{equation}
A \emph{compensated} aperture, fulfilling
\begin{equation}
0=\int d\theta \; \theta \; u(\theta) \; ,
\label{eqn:compensated}
\end{equation}
is insensitive to mass sheets. The compensated aperture mass can be expressed equivalently in terms of tangential shears $\gamma_t$, which are related to $\kappa$ as \citep[see][Eqn.~24, for this azimuthally averaged relation]{saasfee}
\begin{equation}
\gamma_t(\theta) = \langle \kappa(\theta') \rangle_{(\theta'<\theta)} - \kappa(\theta) \; .
\label{eqn:gammatkappa}
\end{equation}
With a corresponding shear filter $q_{\gamma}(\theta)$, 
\begin{equation}
 \Map = 2\pi \int d\theta \; \theta \; q_{\gamma}(\theta) \gamma_t(\theta) \; .
\end{equation}
The tangential shear weight function $q_{\gamma}$ can be calculated from the projected mass density weight function $u$ via
\begin{equation}
q_{\gamma}(\theta)=\frac{2}{\theta^2} \int_0^\theta d\theta' \; \theta' \; u(\theta') - u(\theta) \; 
\end{equation}
and vice versa, using Eqn.~\ref{eqn:gammatkappa}, as
\begin{equation}
u(\theta)=2\int_\theta^{\infty} d\theta'\;\frac{q_{\gamma}(\theta')}{\theta'} - q_{\gamma}(\theta) \; ,
\label{eqn:transformgk}
\end{equation}
which is by definition compensated.

Since the observable is the reduced shear $\bmath{g}=\bgamma/(1-\kappa)$, $\Map$ must in practice be determined as
\begin{equation}
 \Map = 2\pi \int d\theta \; \theta \; q(\theta) g(\theta)
\label{eqn:mapq}
\end{equation}
with reduced shear weights $q(\theta)$. 

We will assume that the integral~\ref{eqn:mapq} will be executed as a sum over measured average tangential reduced shears $g_j$ in a finite number of annuli, weighted as
\begin{equation}
 \Map=\sum_{j=1}^N Q_j g_j = \bmath{g}^T\bmath{Q}\; ,
\end{equation}
where $Q_j$ is the weight for the average tangential reduced shear $g_j$ in annulus $j$. We wish to find the $Q_j$ which minimize the mean squared error of the lensing mass estimator relative to $M_{200m},$ the mass enclosed in a sphere of 200 times the mean matter density:
\begin{equation}
\sigma^2_M \equiv \left\langle (\Map - M_{200m})^2\right\rangle.
\end{equation}

Most previous optimizations have assumed that the shape of the mass profile is known {\it a priori}, such that $g_j = M_{200m} g_{\mathrm{true},j} + \epsilon_j$, where $\epsilon_j$ is a measurement noise with zero mean and known variance.  In this case, the filter $\bmath{Q}$ that minimizes $\sigma^2_M$ is the same as the (Wiener) filter that maximizes signal-to-noise ratio of $\Map$.  When the $\epsilon_j$ are uncorrelated between different annuli (as for shape noise), the Wiener filter is \citep[e.g.][]{2006PhRvD..73l3525M}
\begin{equation}
Q_j\propto g_{\mathrm{true},j} / {\mathrm Var}(\epsilon_j).
\label{eqn:shapenoise}
\end{equation}
Such filters can be constructed analytically for widely used dark matter halo mass profiles, such as the NFW profile \citep{1996ApJ...462..563N}, where the shear is given by \citet{1996A&A...313..697B} and \citet{2000ApJ...534...34W}.   The filter derived under these assumptions---a cluster shear profile assumed to follow the NFW form, with known concentration, and measurement noise due entirely to shape noise---we will henceforth denote as the ``S filter.''

If the measurement noise has non-zero covariance between annuli, as is the case if we consider shear from uncorrelated LSS to be part of the noise, then we can define a covariance matrix via $C_{jk} = \langle \epsilon_j \epsilon_k \rangle$, and we can generalize equation~(\ref{eqn:shapenoise}) to
\begin{equation}
\bmath{Q}\propto \hat{C}^{-1}\bmath{g_{\mathrm{true}}} \; .
\label{eqn:qwiener}
\end{equation}
We will denote as the ``$S+U$'' filter the $\bmath{Q}$ derived with these assumptions of known NFW shear profile with measurement noise from the combination of shape noise plus uncorrelated projected large-scale structures.

Equation~(\ref{eqn:qwiener}) is not optimal, however, when there is variability in the shear profiles of halos. Consider the reduced shear signal of simulated cluster $i$, $1\leq i \leq n$ in annulus $j$, $1\leq j\leq N$ to be given as the element $g_{ij}$ of a matrix $\hat{g}$. Since the simulation does not include shape noise or LSS outside a certain integration region, we assume that measurements will incur an additional noise $\epsilon_j$ with covariance matrix $\hat{C}$ that is independent of the cluster properties. For filter $\bmath{Q}$ the mass estimation variance becomes
\begin{equation}
n\sigma^2_M = (\bmath{M}-\hat{g}\bmath{Q})^{T}(\bmath{M}-\hat{g}\bmath{Q}) + n\bmath{Q}^{T}\hat{C}\bmath{Q},
\label{eqn:mvar}
\end{equation} 
where $\bmath{M}$ is the vector of true masses $M_{200m}$. The minimum variance weights can be found from Eqn.~\ref{eqn:mvar} as
\begin{equation}
\bmath{Q_0} = (\hat{g}^{T}\hat{g}+n\hat{C})^{-1} \hat{g}^{T}\bmath{M} \; .
\label{eqn:qopt}
\end{equation}
These minimum variance weights are not necessarily unbiased, but typically systematically underestimate the mass slightly. We rather minimize the variance under the constraint of no bias, i.e. $\langle \bmath{M}_i \rangle - \bmath{Q}\cdot\langle\bmath{g}\rangle=0$, where $\langle\bmath{g}\rangle$ is the mean reduced shear of the sample, which we add to Eqn.~\ref{eqn:mvar} with a Lagrange multiplier. The resulting minimum variance filter among unbiased linear filters is
\begin{equation}
\bmath{Q}=\bmath{Q_0}+\bmath{\Delta Q}\cdot\frac{\langle \bmath{M}_i \rangle-\langle(\hat{g}\bmath{Q_0})_i\rangle}{\langle(\hat{g}\bmath{\Delta Q})_i\rangle} 
\end{equation}
where $\bmath{\Delta Q}=(\hat{g}^{T}\hat{g}+n\hat{C})^{-1}\langle\bmath{g}\rangle$.

This filter will be denoted as the ``$S+U+C$'' or the ``optimal linear'' filter, since it incorporates correlated structures, and has the lowest possible $\sigma^2_M$ of any linear operation applied to the azimuthally averaged shear profiles of realistic simulated halos.  

In the remainder of the paper we will examine the properties and performance of the optimal linear filter, and the degradation that correlated structure induces on the simpler $S$ and $S+U$ filters.  Note that all three of the filters discussed here depend upon the value of $M_{200m}$ for the cluster: the scale radius of the NFW profile assumed in the $S$ and $S+U$ filter derivations is typically mass-dependent; and the $\hat{g}$ matrix in the $S+U+C$ filter also depends upon the mass range of clusters taken from the simulation for the optimization.  In practice, of course, the cluster mass is not known in advance, and some kind of iterative procedure would be needed to choose a filter that is (approximately) optimized for each cluster's mass.  We will leave treatment of this iterative procedure to future work, and in this paper simplify the task by assigning the simulated clusters to bins based on their known $M_{200m}$ values, and apply a single filter to each cluster-mass bin. Also, we only optimize filters for one redshift of $z_d=0.245$. While the method we develop can be used as a general prescription for optimizing mass measurements, the change in shear amplitudes, angular scales and correlated structure of clusters requires that the filters be recalibrated for each individual redshift.

\subsection[]{Uncorrelated Large Scale Structure}

The presence of large scale structure (LSS) between sources and observer that is uncorrelated to the observed cluster is an additional source of noise that must be accounted for when finding optimal filters for weak lensing measurements \citep{2004PhRvD..70b3008D,2005A&A...442..851M}. We use a halo model approach to calculate the lensing covariance matrix of LSS that is outside the $\pm200\Mpch$ region integrated in the simulation.  We assume that structures beyond $200\Mpch$ are uncorrelated with the target cluster. The halo model also allows to exclude shear noise from uncorrelated halos with $M_{200m}>10^{14}\Msol$. The rationale for this is that if these massive halos were present along the line of sight to a target cluster, they would be detected independently, and the interference could be modelled or the target cluster could be discarded from the sample. For details of the calculation, we refer the reader to Appendix~\ref{appendix:hm}.

\section[]{Results}

\subsection[]{Variability of Halo Profiles}
\label{sec:ipv}
We begin by investigating the mean and variance of halo shear profiles. Taking the average of the simulated cluster shear profiles, we find that both globally and for narrow mass bins it is possible to fit a NFW profile to the cluster with deviations from the average shear that are typically within 5-10\% of the signal. The individual profiles, however, show a great variety of structures. These must be interpreted as due to variations in concentration, halo ellipticity, and mass in the form of substructure and correlated second halos.

\begin{figure}
\centering
\includegraphics[width=0.48\textwidth]{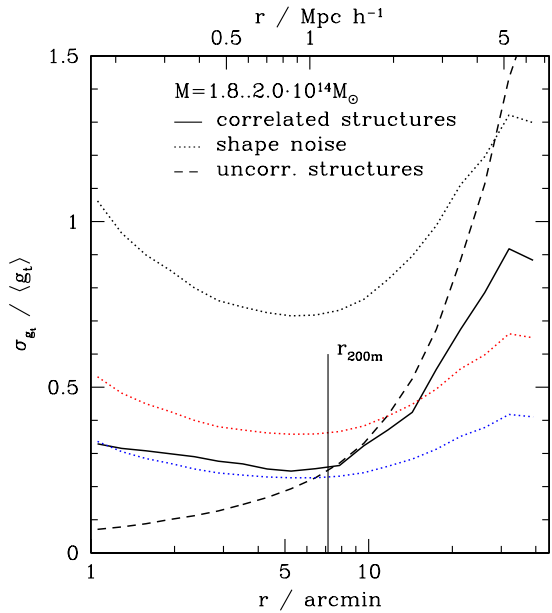}
\caption{Relative shear dispersion due to shape noise of three different levels (dotted lines for 10, 40 and 100 background galaxies per arcmin$^2$), uncorrelated LSS along the line of sight between $z_s=1$ and the observer (dashed) and intrinsic profile variance due to correlated structures (solid line). The latter is calculated for halos with a narrow range of masses ($1.8\ldots2.0\cdot10^{14}\Msol$) with respect to their mean shear signal.}
\label{fig:sigmagammaogamma}
\end{figure}

In narrow mass bins, we find the covariance matrix of the shear signal in our radial bins with respect to the empirical mean. We estimate the expected covariance matrix from uncorrelated LSS within the $\pm200\Mpch$ integration region of the simulations using the halo model approach described in Appendix~\ref{appendix:hm}, and subtract this from the covariance ``observed'' in the simulation to yield a matrix that can be fully attributed to internal halo variation and correlated halos. Figure~\ref{fig:sigmagammaogamma} shows the dispersion of shears in relation to the mean signal as a function of radius for an example mass bin. The overplotted dispersions due to shape noise of different levels and uncorrelated LSS between background sources at $z=1$ and observer show that the intrinsic variance has its highest relative importance inside and at the virial radius, 2--7\arcmin, inside of which the shape noise and outside of which the noise of uncorrelated LSS increase more steeply than the intrinsic variance. In this range, the intrinsic dispersion of the profile is close to proportional to the mean shear signal. Note that the intrinsic variance and LSS are correlated between radial bins, unlike the shape noise, so this plot somewhat over-represents the relative impact of shape noise on mass estimation.

\begin{figure}
\centering
\includegraphics[width=0.48\textwidth]{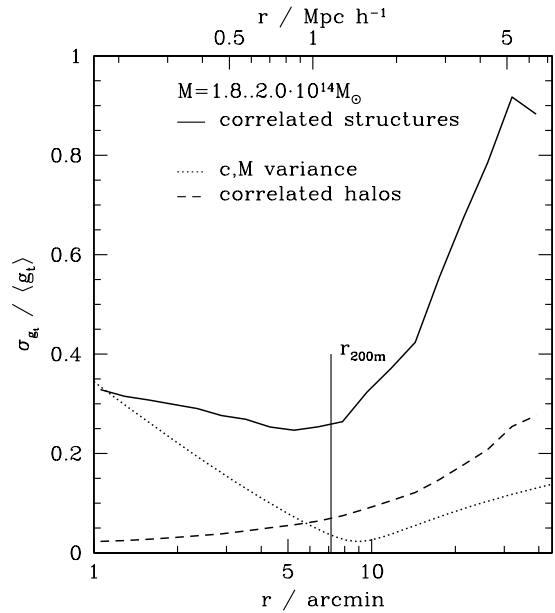}
\caption{Relative variations of halo shear profiles (solid line) in a narrow range of masses ($1.8\ldots2.0\cdot10^{14}\Msol$) and contributions to this from the expected distribution of profiles from variations in concentration with $\sigma_{\log_{10} c_{200m}}=0.18$ and masses (dotted) and Poisson noise in excess density of second halos due to linear two-point correlation (dashed). The variance of shear profiles has contributions from uncorrelated LSS within the integration region of the simulations subtracted.}
\label{fig:sigmagammaogammacorr}
\end{figure}

Figure~\ref{fig:sigmagammaogammacorr} plots the intrinsic dispersion of shears due to correlated structures in relation to the mean signal along with some expected contributions to this. The dotted line plots the dispersion in $g_t$ expected from the variance of concentration values for the best-fit NFW profiles to the halos.  We follow \citet{2001MNRAS.321..559B} and take a log-normal distribution of concentrations with $\sigma_{\log_{10} c_{200m}}=0.18$. The dotted line also includes the range of $M_{200m}$ variation within the bin, but because the mass bin is narrow, the variation of concentrations dominates this component. The concentration variation explains most of the shear variance at $r\approx1'$ \citep[in agreement with][]{2010arXiv1008.1579R}, but is subdominant at larger radii. The overdensity of correlated neighboring halos introduces additional variance, which we calculate using a halo model by assuming a Poisson distribution of mutually independent halos (see Appendix). This neighbor-halo shot noise becomes more important towards the outskirts. These two components, however, are not enough to explain the entire variability of shear signals. Additional contributions that are difficult to estimate analytically stem from the aspherical halo profile, the distribution of neighboring halos, and substructure of halos, for instance due to recent mergers. Simulations remain the most viable way of assessing these.

\subsection[]{Improvements in Mass Uncertainty}
\label{sec:imu}
We compare here the variance in halo masses obtained with three different linear filters $\bmath{Q}$ for the tangential shear profile:
\begin{enumerate}
\item The $S$ filter, which is the Wiener optimal filter for measuring halos that are assumed to have the mean (NFW) reduced shear profile of the individual mass bin, with uncertainty from shape noise only, as per Equation~(\ref{eqn:shapenoise});
\item The $S+U$ Wiener filter that in addition accounts for shear variance from uncorrelated LSS, as per Equation~(\ref{eqn:qwiener}), and 
\item Our minimum-variance $S+U+C$ filter that takes into account the full variance including correlated structures and variation of internal halo structure, as per Equation~(\ref{eqn:qopt}).
\end{enumerate}
Although these filters were derived under different assumptions about sources of shear variance, they are always {\em evaluated} taking into account the full noise in the lensing signal due to shape noise, plus uncorrelated LSS, plus the correlated structure as measured in the $N$-body simulations.  We consider shape noise levels corresponding to an intrinsic ellipticity dispersion of $\sigma_\varepsilon=0.3$ and three survey depths of 10, 40 and 100 background galaxies per arcmin$^2$. 
We calculate the uncertainty of mass estimates based on the different filters as
\begin{equation}
\sigma_M/M = \sqrt{\langle(\bmath{Q}\cdot\bmath{g_i}-\bmath{M}_i)^2\rangle+\bmath{Q}^T \hat{C} \bmath{Q}}/\langle \bmath{M}_i\rangle \;,
\label{eqn:sigmamom}
\end{equation}
where $\bmath{g_i}$ is the vector of radially binned tangential reduced shears for halo $i$ extracted from the simulations without shape noise and without LSS outside the integration region, $\bmath{M}_i$ is the true $M_{200m}$ mass of halo $i$ as measured from the simulations and $\hat{C}$ is the covariance matrix due to shape noise and uncorrelated LSS outside the integration region. In order to find the dependence on cluster mass we split the sample into narrow mass bins with a population of at least 700 halos each and find $\sigma_M/M$ for each mass bin at three different levels of shape noise. 

\begin{figure}
\centering
\includegraphics[width=0.36\textwidth]{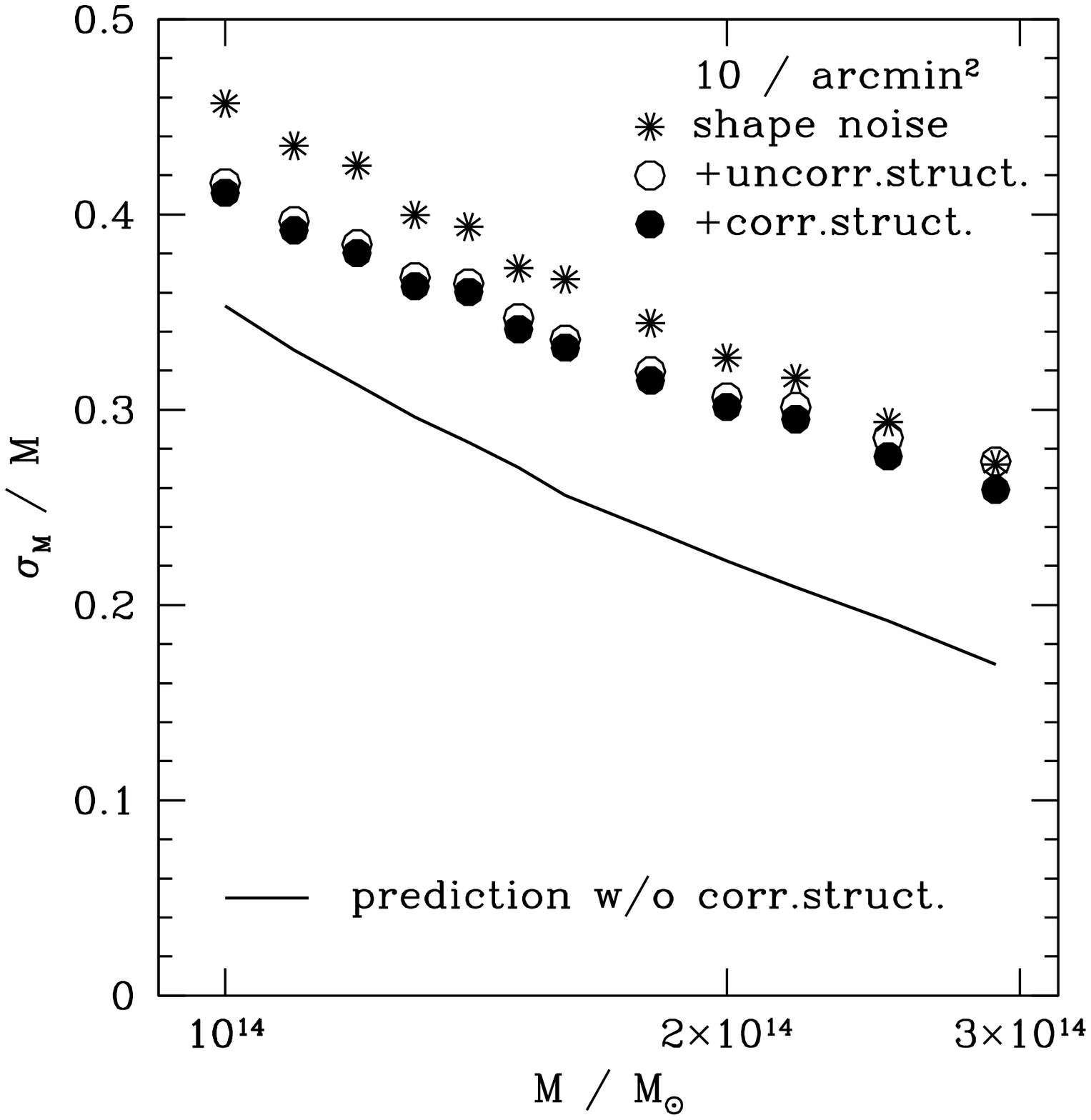}
\includegraphics[width=0.36\textwidth]{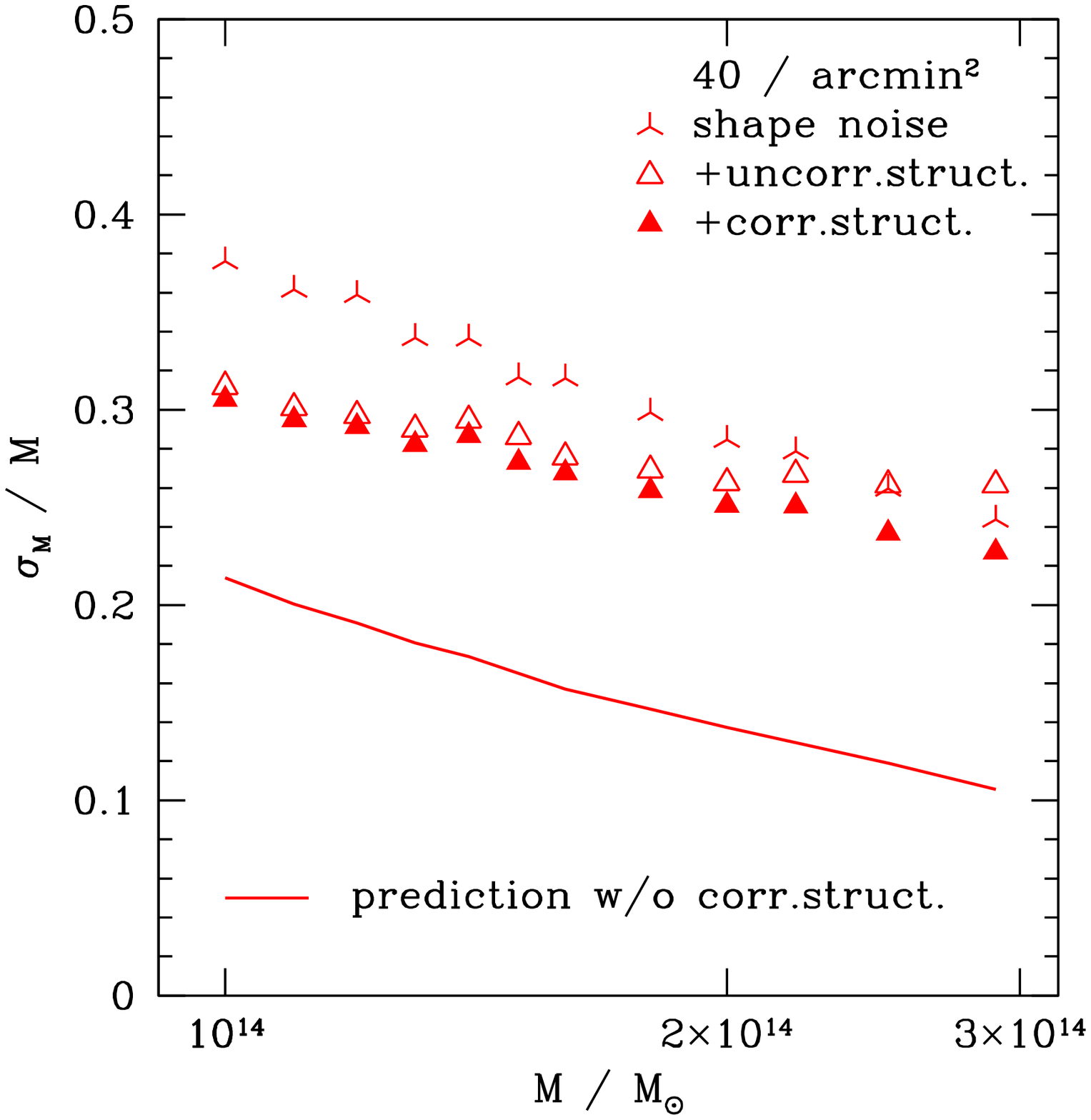}
\includegraphics[width=0.36\textwidth]{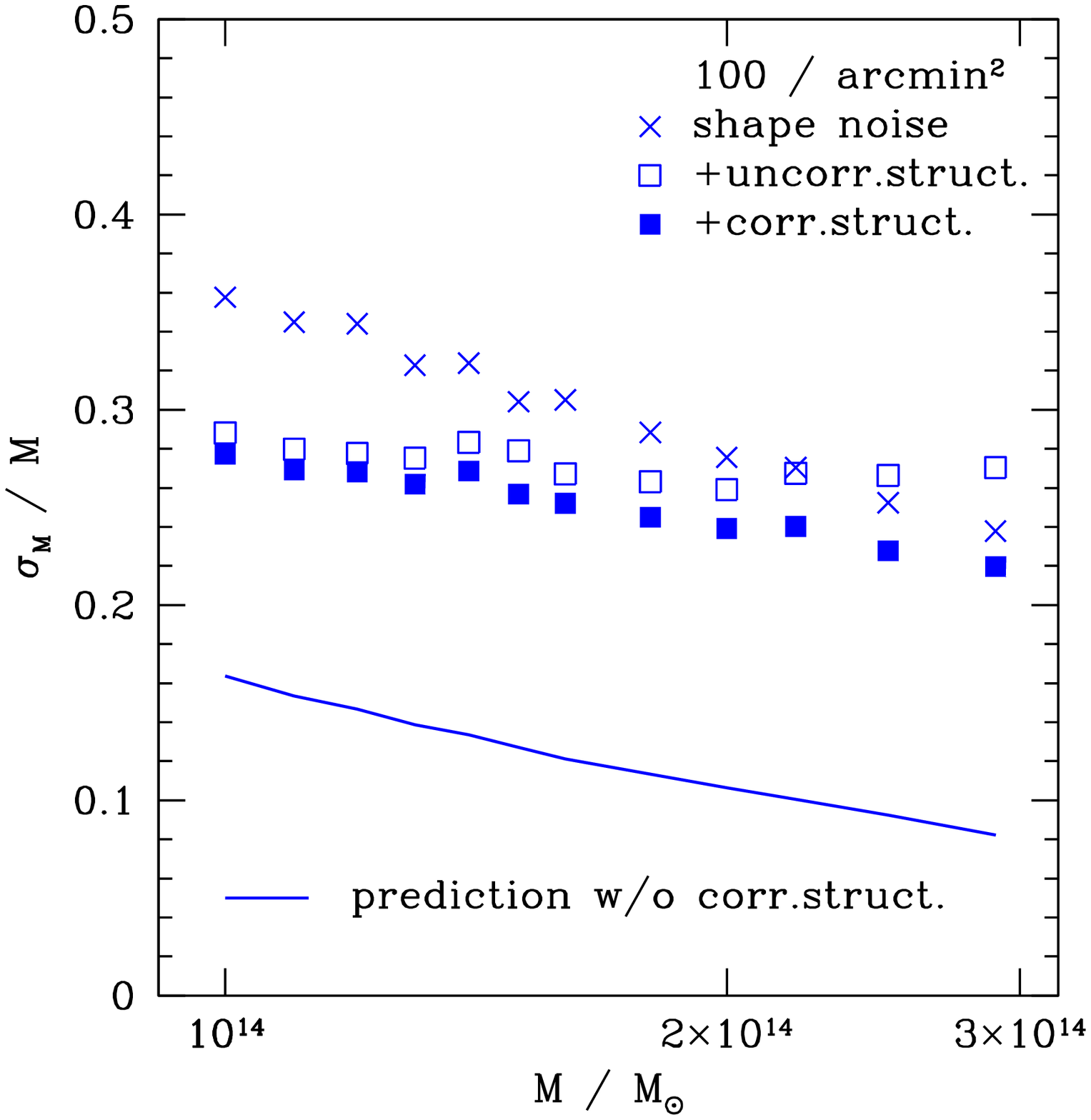}
\caption{
Relative errors in estimates of $M_{200m}$ for different aperture mass shear filters, when applied to simulated $N$-body halos at $z=0.245$.  All symbols show measurement errors including shape noise, uncorrelated LSS, halo profile variation, and correlated structures. Each panel shows RMS errors from: the $S$ filter, optimized for NFW clusters with shape noise only (skeletal symbols); $S+U$ filters optimized for NFW clusters accounting for uncorrelated LSS as well as shape noise (open); and the $S+U+C$ minimum variance filters (solid symbols) optimized with the inclusion of halo profile variation and correlated structures. Also plotted is the prediction for $S+U$ filters in the unrealistic case that all halos have the same profile with no correlated structures (solid line).
}
\label{fig:massdispersion}
\end{figure}

Figures~\ref{fig:massdispersion} plot $\sigma_M/M$ from Eqn.~\ref{eqn:sigmamom} versus $M_{200m}$ for the three different filters at each survey depth. We further compare the result to the naive prediction 
\begin{equation}
\sigma_M/M=\sqrt{\bmath{Q}^T \hat{C} \bmath{Q}}/\langle \bmath{M}_i\rangle
\label{eqn:naive}
\end{equation}
where $\bmath{Q}$ is the $S+U$ filter optimized for the mean shear profile of the bin in the presence of uncorrelated LSS and shape noise.
This is the mass uncertainty one would expect if all clusters were true NFW halos with no scatter in concentration $c(M,z)$ at constant mass and redshift, and no correlated structure was present. Statistical $1\sigma$ errors in $\sigma_M/M$ due to the limited size of our sample of halos
were estimated with a jackknife approach (i.e. by leaving out one halo from the analysis in turns) and are always below $0.01$.

The following results are of interest:
\begin{itemize}
\item The $S+U+C$ filter has the lowest mass estimate variance, which is true by definition. 
\item The uncertainty of the mass estimates with the $S+U$ filter predicted for the naive case of no correlated structure is always significantly lower than any real uncertainty. 
\item Due to variance in the intrinsic profile, which can be mitigated neither by deeper data nor by heavier clusters, there is a lower limit to mass uncertainty. Deeper surveys benefit mass estimates remarkably little, especially for relatively massive halos. Observations beyond a background galaxy density of 40~arcmin$^{-2}$ are not notably improving mass estimates for the mass range of halos and redshift $z_d=0.245$ considered here, and will only benefit cases where shape noise is more dominant, i.e. less massive clusters and objects at higher redshift. 
\item On the high signal-to-noise end of massive halos and low shape noise, the $S+U$ filter that is optimized for uncorrelated LSS and shape noise, but not for intrinsic variability of profiles, yields a mass estimate that is no improvement over the $S$ filter without taking LSS into account. 
\item On realistic ground-based data with a background galaxy density of 10~arcmin$^2$, the improvement from taking into account intrinsic profile variability is very small since the $S+U$ and $S+U+C$ filters are almost identical and the large shape noise in the center prevents the $S+U$ filter from assigning too much weight to the core. The increase in mass uncertainty from the naive prediction, however, remains relevant even at these observational conditions.
\end{itemize}

Can a non-linear fitting process improve upon our optimal linear shear estimates of $M_{200m}$?
For comparison, we note that for two-dimensional fitting of an NFW profile with $M_{200m}$ and $c_{200m}$ as free parameters, a sample with halos of mass $M_{200m}>4\cdot10^{14}\Msol$ from the same simulation has $\sigma_M/M=0.4$ ($0.27$) for shape noise levels of 10 (40)~arcmin$^{-2}$ \citep{2010arXiv1011.1681B}. We do not have results for the exact corresponding mass range and the methods differ in that we pre-sort halos into true mass bins. However, the comparison of mass uncertainties suggests that the two-parameter non-linear fit does not yield significant improvements. This is potentially because the contribution of concentration variation to the intrinsic variability of profiles is subdominant (cf. Section~\ref{sec:ipv}) and a degeneracy between mass and concentration exists.

\subsection[]{A Look at Optimized Filters}
\label{sec:laf}
\begin{figure}
\centering
\includegraphics[width=0.48\textwidth]{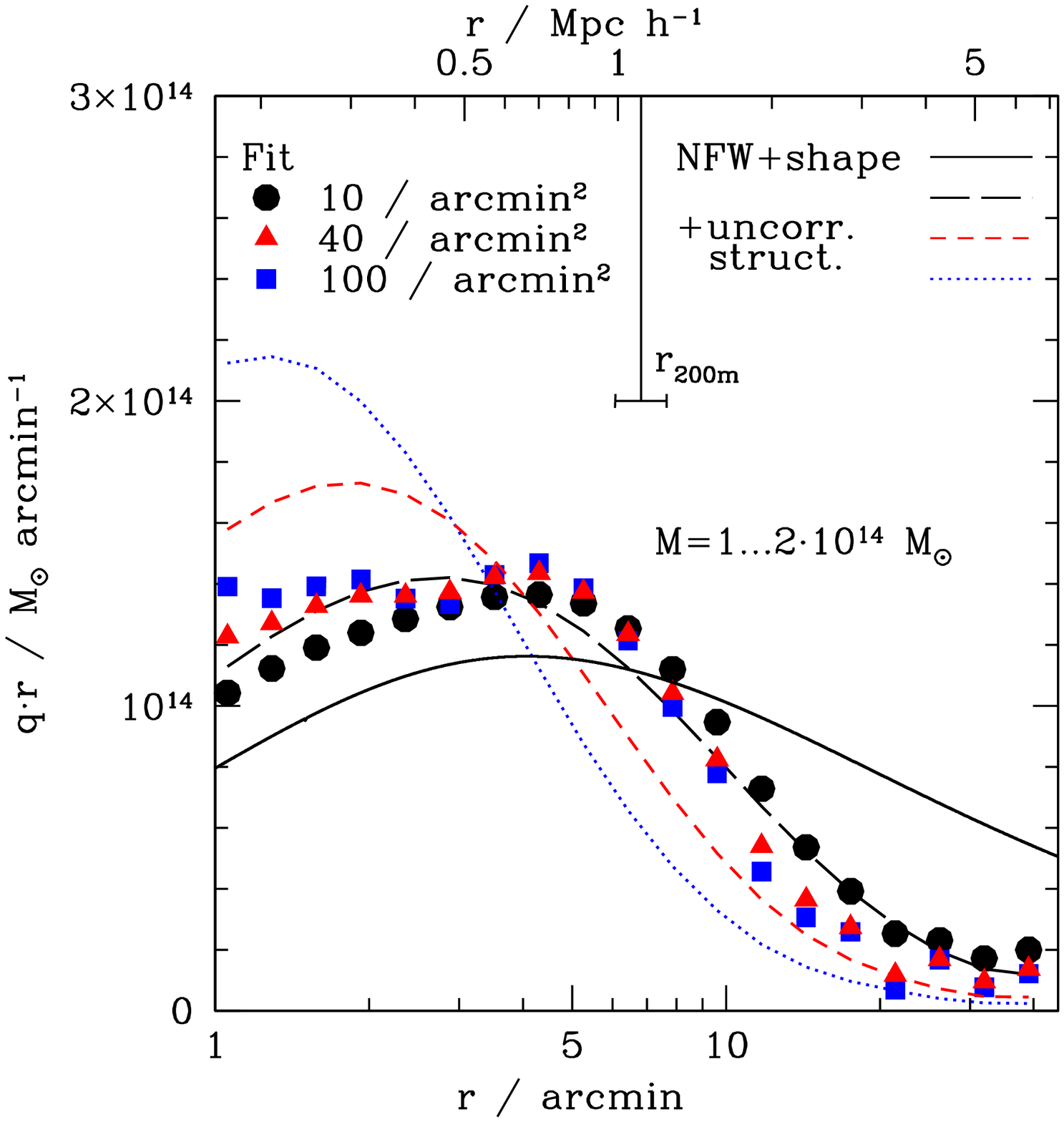}
\includegraphics[width=0.48\textwidth]{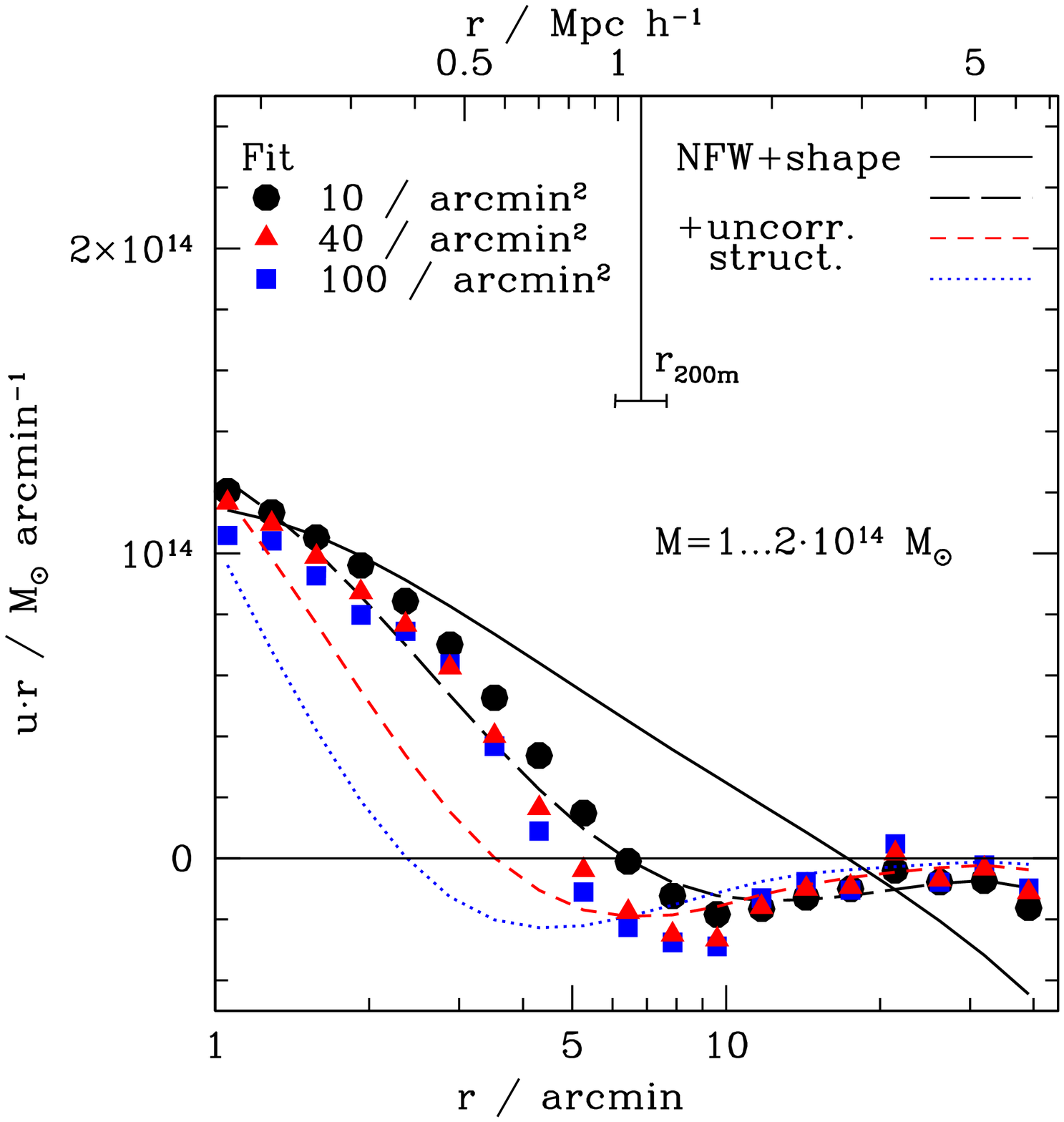}
\caption{Fitted weights for different levels of shape noise (circular, triangular and square symbols for 10, 40 and 100 background galaxies per arcmin$^2$, respectively). Plotted for comparison are NFW Wiener filters for shape noise only (solid line) and for LSS and the three levels of shape noise (long dashed, short dashed and dotted lines). The two panels show tangential shear weights and convergence weights transformed according to the transformation of reduced shear weights $q(\theta)$ into $q_{\gamma}(\theta)\approx q(\theta)/(1-\kappa(\theta))$ and application of Eqn.~\ref{eqn:transformgk}.}
\label{fig:weights}
\end{figure}

We briefly examine the optimal linear filters found by our procedure. These take into account the covariance due to uncorrelated LSS along the line of sight and shape noise while minimizing the variance of the mass estimate empirically by using the simulated cluster shear profiles. Figure~\ref{fig:weights} compares these $S+U+C$ weights to the $S$ and $S+U$ filters optimized for NFW-profile clusters.

The optimized filters differ from the NFW-optimized filters in that they do not put as much weight to the innermost region, shifting the maximum of $q\cdot r$ to $r\approx5'$ for the mass bin $1\ldots2\cdot10^{14}\Msol$. In terms of weights on convergence $u$, this puts more weight on projected density at intermediate distance from the core. Consequently, the convergence filter $u$ compensates at larger radii. All filters that take LSS into account, however, do put less weight on the shear signal outside $r\approx10'$, where the LSS signal becomes comparable to or larger than shape noise.

This suggests an explanation for why the $S+U$ filter that includes uncorrelated LSS can perform worse than the $S$ filter optimized only for shape noise:
the $S+U$ filter places higher weight on the intrinsically variable central region, generating noise that outweighs the gain from putting less weight to the outer regions where uncorrelated LSS dominates the errors.

\citet{2007A&A...462..875S}, \citet{2010MNRAS.405.2078M} and \citet{2010arXiv1011.1084H} discuss various reasons for the high uncertainty of shear measurements near the core and the benefits of downweighting the shear signal in that region. Some of the effects mentioned are not present in our simulations, such as baryonic effects, signal dilution due to cluster member galaxies (which are most prominent near the center), and magnification-induced change and incomplete sampling of the redshift distribution of sources when no or only uncertain single source redshift estimates are present. In practical application, this would degrade the performance of $S+U$ filters even more since they weight the central regions more heavily.

\section[]{Summary}

By the linear least squares method outlined in Section~\ref{sec:mvf} we have found filters for the tangential shear signal of simulated halos that minimize the variance between mass estimates and true $M_{200m}$ of the clusters in the presence of shape noise, uncorrelated projected LSS and the intrinsic variability of halo shear profiles. We have shown in Section~\ref{sec:imu} that taking into account the last source of noise can improve upon the mass estimates of filters that assume a constant NFW profile. This improvement is comparable to the improvement that has been previously noted as resulting from the inclusion of uncorrelated LSS in the filter optimization, compared to optimizing solely for shape noise at $M\geq1.5\cdot10^{14}\Msol$ and a background galaxy density of $40$arcmin$^{-2}$ or more. The most obvious difference from filters that do not take intrinsic profile variability into account is a suppression of weight on the central region of the halo (cf. Section~\ref{sec:laf}), where intrinsic variability of profiles has its most important contribution to the overall noise (cf. Section~\ref{sec:ipv}).

Apart from these observations, this has a number of important conclusions for weak lensing measurements of cluster masses:
\begin{itemize}
\item Even when accounting for shape noise and noise due to uncorrelated LSS, the uncertainties of an aperture mass measurement will be substantially underestimated if we neglect the intrinsic variability of halo profiles. The intrinsic variability of halo shear profiles sets a lower limit to the accuracy of lensing mass estimations which, especially for low shape noise and relatively massive halos, can be a multiple of what would be expected for fixed halo profiles. This has to be accounted for when calibrating the MOR, even more so as the errors in mass estimators must be known to great accuracy for cluster count probes to be successful \citep{2005PhRvD..72d3006L}. It also means that, at least with this method of estimating cluster masses with weak lensing, there is limited benefit in making ever deeper observations, because uncertainty in the signal remains both in the core (because of intrinsic variability) and in the outer parts (because of LSS) of halos. For more massive halos, the problem persists since at least some components of the intrinsic variability of the profile (e.g. concentration distribution, asphericity, substructure) are likely to scale with the overall profile amplitude. Ground based surveys and even more so space-based surveys and deep space-based observations will have to take this into account when predicting the accuracy of their MOR calibration as a function of survey depth and width. Deeper lensing surveys will show advantage only when calibrating clusters at redshifts well away from the $z=0.245$ studied herein.
\item The intrinsic variations in shear profiles can be explained by variations in concentration only in the core regions of halos \citep[cf. also][]{2010arXiv1008.1579R}. It is important to take into account as well the variations of profiles due to substructure and correlated structures, which dominate the intrinsic profile variability near and outside the virial radius.
\item When, in addition to shape noise, aperture mass filters account for uncorrelated LSS only \citep{2004PhRvD..70b3008D,2005A&A...442..851M}, the uncertainty of mass estimates can actually become larger. This is a result of the intrinsic profile variability near the core of the halo, where LSS optimized filters put more weight than would a filter optimized for the influence of both uncorrelated LSS and intrinsic variability. We expect optimization for uncorrelated LSS to be non-beneficial for clusters of mass $M_{200m}\approx3.0\cdot10^{14}\Msol$ at the simulated redshift of $z=0.245$ and background galaxy levels above $10$~arcmin$^{-2}$.
\end{itemize}

\section*{Acknowledgments}

We thank Anatoly Klypin for providing the simulation and Jeremy Tinker for providing the halo finder used in this study.  Additionally, we thank Matthew R. Becker and Andrey Kravtsov for providing us with the weak lensing shear and convergence maps from the simulations. The authors are grateful to Matthew R. Becker for helpful comments and Joseph J. Mohr for discussions on the MOR. TYL would like to thank the hospitality of the University of Pennsylvania where this collaboration started.

DG acknowledges support by Studienstiftung des deutschen Volkes and DAAD (Deutscher Akademischer Austauschdienst). 
GB was supported by NSF grant AST-090827 and DOE grant DE-FG02-95ER40893. TYL was supported by the World Premier International Research Center Initiative (WPI Initiative), MEXT, Japan and a JSPS travel grant (Institutional Program for Young Researcher Overseas Visits). SS acknowledges support by TR33 "The Dark Universe" and the DFG Cluster of Excellence on the "Origin and Structure of the Universe".

\addcontentsline{toc}{chapter}{Bibliography}
\bibliographystyle{mn2e}
\bibliography{literature}

\appendix

\section[]{Halo Model}

\label{appendix:hm}

In the following we describe the halo model used to calculate the covariance of shear in annuli due to structure uncorrelated to the central halo and outside the integration region of the simulations. For a general review of halo models of large scale structure, we refer the reader to \citet{2002PhR...372....1C}. Approaches similar to ours are presented in \citet[][in Fourier space]{2000ApJ...535L...9C} and \citet[][in real space]{2003MNRAS.344..857T}.

Consider a set of halos $1,\ldots,i,\ldots,n$ with $0<z_i<z_s=1$. The tangential shear $\gamma_t^k$ they introduce in an annulus $k$ is
\begin{equation}
\gamma_t^k = \sum_i \gamma_t^k(\bmath{h_i}) \; ,
\label{eqn:gtk}
\end{equation}
where $\bmath{h_i}=(\theta_{1,i},\theta_{2,i},z_i,M_i,c_{i},\ldots)$ is a set of parameters describing the position and mass profile of halo $i$ and $\gamma_t^k(\bmath{h_i})$ is the tangential shear such a halo introduces on annulus $k$.

We can then straightforwardly calculate the covariance of $\gamma_t^k$ and $\gamma_t^l$ in annuli $k$ and $l$ as
\begin{eqnarray}
\left\langle \gamma_t^k \cdot \gamma_t^l \right\rangle = \left\langle\sum_i \gamma_t^k(\bmath{h_i})\cdot\sum_j \gamma_t^l(\bmath{h_j})\right\rangle = \nonumber \\
\left\langle\sum_i\left(\gamma_t^k(\bmath{h_i})\cdot \gamma_t^l(\bmath{h_i})\right)\right\rangle + \left\langle\sum_{i\neq j} \left(\gamma_t^k(\bmath{h_i})\cdot \gamma_t^l(\bmath{h_j})\right)\right\rangle \; ,
\label{eqn:gtcov1}
\end{eqnarray}
where the average goes over random realizations of the halo populations. Note that we have separated the terms caused by the shear of a single halo on both annuli (one-halo term) from correlations caused by pairs of halos (two-halo term). The latter will not be zero, because halo positions are correlated.

The average in Eqn.~\ref{eqn:gtcov1} can be expressed as an integral over probabilities $dP_1(\bmath{h})$ of finding a halo with properties $\bmath{h}$ and $dP_2(\bmath{h_1},\bmath{h_2})$ of simultaneously finding two halos with properties $\bmath{h_1}$ and $\bmath{h_2}$, i.e.
\begin{eqnarray}
\mathrm{Cov}(\gamma_t^k,\gamma_t^l) = \int dP_1(\bmath{h}) \; \gamma_t^k(\bmath{h}) \cdot \gamma_t^l(\bmath{h}) \nonumber \\ + \int dP_2(\bmath{h_1},\bmath{h_2}) \; \gamma_t^k(\bmath{h_1}) \cdot \gamma_t^l(\bmath{h_2}) \; .
\label{eqn:covint}
\end{eqnarray}

The probability density $dP_1(\bmath{h})$ of finding a halo with properties $\bmath{h}$ has as its cosmological ingredients the halo mass function,
$\frac{dn}{dM}(M,z)$, and the distribution of halo density profiles. We can write
\begin{equation}
dP_1(\bmath{h}) = \frac{dn}{dM}\; dM\; \frac{dV}{d\Omega\;dz}\; d\Omega\; dz \; p(c|M,z)\; dc \; ,
\label{eqn:dp11}
\end{equation}
assuming NFW halo profiles and a probability density $p(c|M,z)$ of finding concentration parameters $c\equiv c_{200m}=r_{200m}/r_s$ which only depends on the halo mass $M\equiv M_{200m}$ and redshift $z$. Note that we use definitions of $r_{200m}$, $c_{200m}$, and $M_{200m}$ with respect to 200 times the mean matter density at the halo's redshift. We assume as a fixed mass-concentration relation the best-fit formula from \citet{2008MNRAS.390L..64D} \citep[using WMAP5 cosmology, cf.][]{2009ApJS..180..330K},
\begin{equation}
c(M,z)=10.14\cdot\left(\frac{M}{2\cdot10^{12}\Msol}\right)^{-0.081}\cdot(1+z)^{-1.01}\; .
\label{eqn:c200mz}
\end{equation}
For the halo mass function and the mass dependent biases $b(M)$ used below, we adopt the Sheth-Tormen mass function and the associated bias \citep{1999MNRAS.308..119S}.

The combined probability density $dP_2(\bmath{h_1},\bmath{h_2})$ of simultaneously finding two halos with properties $\bmath{h_1}$ and $\bmath{h_2}$ must include the interdependence of their probabilities. Assume that halos cluster according to a 3d two-point correlation function $\xi_{hh}(r,\bmath{h_1},\bmath{h_2})$ which is generally dependent on halo properties $\bmath{h_{1/2}}$ in addition to their separation $r$. We can then write
\begin{eqnarray}
dP_2(\bmath{h_1},\bmath{h_2}) = \nonumber \\
dP_1(\bmath{h_1})\cdot dP_1(\bmath{h_2})\cdot \left(1+\xi_{hh}(\bmath{h_1},\bmath{h_2})\right) \; .
\label{eqn:dp2}
\end{eqnarray}
Because the tangential shear due to a randomly placed halo does not correlate with the shear due to another randomly placed halo, the only contribution to the covariance stems from the term proportional to $\xi_{hh}$, and we can drop the other part for the numerical integration. In our halo model, what contributes to the correlation of halo positions is the \emph{linear} evolution of structures. The non-linear evolution, i.e. the formation of halos themselves, need not be considered to this end, as it is dominant only on scales smaller than the ones considered here. The halo-halo correlation function $\xi_{hh}$ can therefore be written as \citep{2010MNRAS.408..300G}
\begin{equation}
\xi_{hh}(r,M_1,M_2,z)\approx b(M_1,z)\cdot b(M_2,z)\cdot \xi_{\mathrm{lin}}(r,z) \; ,
\end{equation}
giving the excess probability of finding a pair of halos with masses $M_1$ and $M_2$ at comoving separation $r$ at redshift $z\approx z_1\approx z_2$. We use a linear power spectrum that is consistent with WMAP 5 results. We relate this to an angular correlation at redshift $z$ by performing one redshift-space integral over the three-dimensional two-point correlation at a fixed angular separation and setting the redshift of two correlated halos to be equal for all other purposes.

We calculate the shear signal of off-center halos on annuli $\gamma_t^k(\bmath{h})$ by integrating the surface density of the halo inside and on the annulus and applying Eqn.~\ref{eqn:gammatkappa}. We take into account the width $\Delta=0.1$ of our annuli $k$ with $r\in[(1-\Delta)\cdot r_k,(1+\Delta)\cdot r_k]$ by calculating and tabulating
\begin{eqnarray}
\gamma_t^k(\bmath{h})=\langle\gamma_t(x,\bmath{h})\rangle_{(1-\Delta)\cdot r_k<x<(1+\Delta)r_k} = \nonumber \\
\frac{1}{2\Delta r^2} \int_{(1-\Delta)r}^{(1+\Delta)r} R\; dR\; \left( \frac{\int_0^R \kappa(x,\bmath{h}) x dx}{R^2/2} - \kappa(R,\bmath{h}) \right) \; ,
\end{eqnarray}
where $\gamma_t(x,\bmath{h})$ and $\kappa(x,\bmath{h})$ are the average tangential shear and convergence a halo with properties $\bmath{h}$ introduces to an annulus of radius $x$. The integral is numerically not more demanding than the simpler one required for zero-width rings. As a simplification for numerical integration, we truncate the projected surface density of NFW halos \citep{1996A&A...313..697B,2000ApJ...534...34W} at a radius $r_{\mathrm{trunc}}=20\cdot r_s>r_{200m}$, which does not change the signal significantly.

When doing the final integrations over $dz$ to get the covariance matrices, we leave out the range $0.162 < z < 0.338$ as it is already included in the simulations. We limit the mass range of halos included to $10^8\Msol\leq M_{200m}\leq 10^{14}\Msol$. This lowers the overall variance because of the lack of more massive halos, which can be expected to be modelled separately in real-world weak lensing analyses \citep{2010arXiv1011.1084H}.

We apply the approximation in which the uncorrelated LSS noise can simply be added to the reduced shear signal of the clusters. Formally taking into account the effect of multiple lensing planes on reduced shears requires more complex calculations.

\subsection*{Results}

\begin{figure}
\centering
\includegraphics[width=0.48\textwidth]{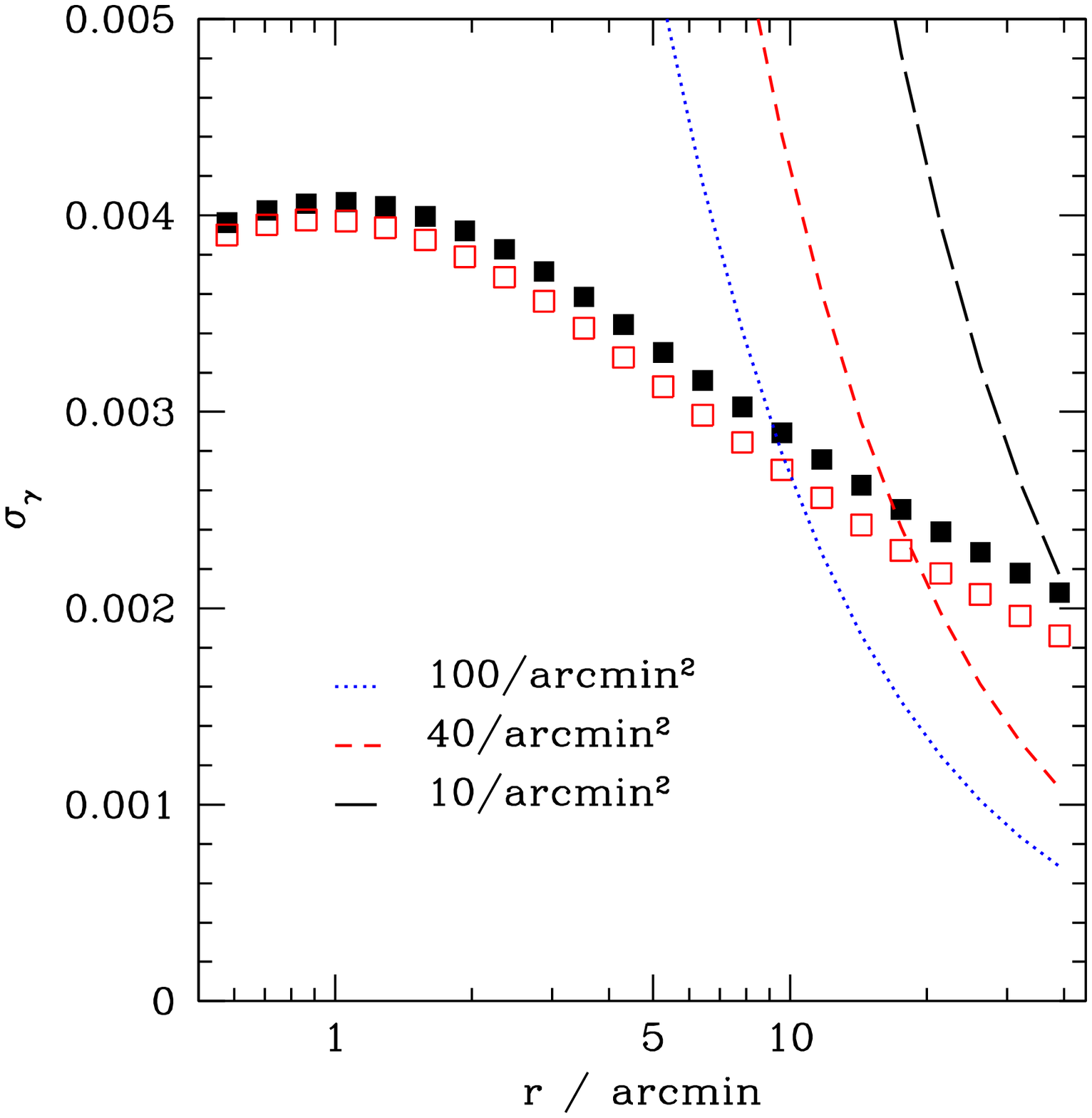}
\includegraphics[width=0.48\textwidth]{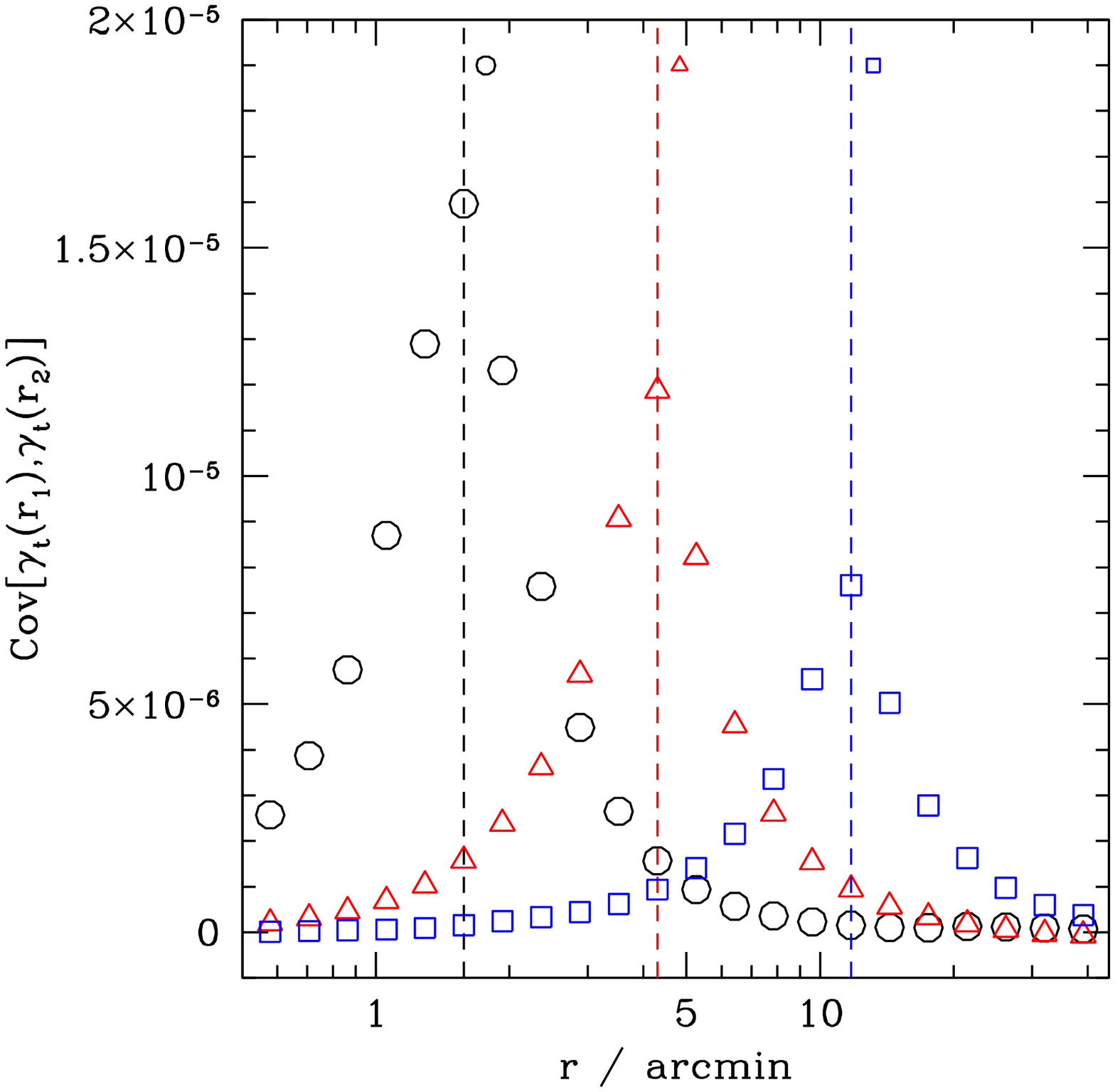}
\caption{Dispersion and covariances of tangential shear signal due to uncorrelated LSS. Dispersions (upper panel) are plotted for the complete redshift range (filled) and when removing contributions from the range of the numerical simulations (open symbols) and compared to different levels of shape noise (lines). Covariances (lower panel) are plotted with respect to three different radii, corresponding to the three different symbols.}
\label{fig:hmvar}
\end{figure}

Shear dispersions as a function of radius compared with shape noise for our binning scheme are shown in Figure~\ref{fig:hmvar}, together with covariances between the signal at different radii. The inner peak is dominated by the one-halo term while the slowly decreasing outer range is mostly due to two-halo contributions. The results from our calculation agree with those presented in \citet{2001A&A...370..743H,2010arXiv1011.1084H}. 

\subsection*{Poisson Covariance due to Correlated Halos}

In a similar manner, we find the variance of the shear signal due to the spherical excess density of second halos around the halo under consideration. The shear signal in annulus $k$ due to this can be expressed as in Eqn.~\ref{eqn:gtk}, where the summation runs over correlated halos. Consequently, the shear covariance due to correlated halos
\begin{equation}
\mathrm{Cov}_c(\gamma_t^k,\gamma_t^l) = \int (dP_1(\bmath{h}|\bmath{h_0})-dP_1(\bmath{h})) \; \gamma_t^k(\bmath{h}) \cdot \gamma_t^l(\bmath{h}) \; ,
\label{eqn:covc}
\end{equation}
can be determined from the excess probability of halos with properties $\bmath{h}$ given the presence of a central halo with properties $\bmath{h_0}$. The term $dP_1(\bmath{h}|\bmath{h_0})-dP_1(\bmath{h})$ is equal to Eqn.~\ref{eqn:dp11} multiplied with $\xi_{hh}(\bmath{h_0},\bmath{h_1})$. This approach makes the assumption that the probabilities of finding two second halos with properties $\bmath{h_1}$ and $\bmath{h_2}$ are mutually independent, which is why the covariance calculated in Eqn.~\ref{eqn:covc} is merely a lower limit to the true shear covariance due to clustering correlated second halos.

\label{lastpage}

\end{document}